\documentclass[%
reprint,
superscriptaddress,
nofootinbib,
amsmath,amssymb,
aps,
prl,
]{revtex4-2}

\usepackage[T1]{fontenc} % Needed for some author names
\usepackage{graphicx}% Include figure files
\usepackage{dcolumn}% Align table columns on decimal point
\usepackage{bm}
\usepackage{hyperref}
\usepackage[all]{hypcap}
\usepackage{amsmath}
\usepackage{mathtools}
\usepackage{tabularx}
\usepackage{multirow}
\usepackage{siunitx}
\usepackage{xcolor}

\begin{document}

\setcounter{secnumdepth}{1}

\title{Search for Gamma-ray Spectral Lines from Dark Matter Annihilation up to 100 TeV towards the Galactic Center with MAGIC}

% author 13.04.2022 Format PRL
\author{H.~Abe}
\affiliation{Japanese MAGIC Group: Institute for Cosmic Ray Research (ICRR), The University of Tokyo, Kashiwa, 277-8582 Chiba, Japan}
\author{S.~Abe}
\affiliation{Japanese MAGIC Group: Institute for Cosmic Ray Research (ICRR), The University of Tokyo, Kashiwa, 277-8582 Chiba, Japan}
\author{V.~A.~Acciari}
\affiliation{Instituto de Astrof\'isica de Canarias and Dpto. de  Astrof\'isica, Universidad de La Laguna, E-38200, La Laguna, Tenerife, Spain}
\author{T.~Aniello}
\affiliation{National Institute for Astrophysics (INAF), I-00136 Rome, Italy}
\author{S.~Ansoldi}
\affiliation{Universit\`a di Udine and INFN Trieste, I-33100 Udine, Italy}\affiliation{also at International Center for Relativistic Astrophysics (ICRA), Rome, Italy}
\author{L.~A.~Antonelli}
\affiliation{National Institute for Astrophysics (INAF), I-00136 Rome, Italy}
\author{A.~Arbet Engels}
\affiliation{Max-Planck-Institut f\"ur Physik, D-80805 M\"unchen, Germany}
\author{C.~Arcaro}
\affiliation{Universit\`a di Padova and INFN, I-35131 Padova, Italy}
\author{M.~Artero}
\affiliation{Institut de F\'isica d'Altes Energies (IFAE), The Barcelona Institute of Science and Technology (BIST), E-08193 Bellaterra (Barcelona), Spain}
\author{K.~Asano}
\affiliation{Japanese MAGIC Group: Institute for Cosmic Ray Research (ICRR), The University of Tokyo, Kashiwa, 277-8582 Chiba, Japan}
% ...

\collaboration{MAGIC Collaboration}
\noaffiliation
\author{N. Hiroshima}
\affiliation{Department of Physics, University of Toyama, 3190 Gofuku, Toyama 930-8555, Japan}
\affiliation{RIKEN iTHEMS, Wako, Saitama 351-0198, Japan}
\author{K. Kohri}
\affiliation{Theory Center, IPNS, KEK, Tsukuba, Ibaraki 305-0801, Japan}
\affiliation{The Graduate University for Advanced Studies (SOKENDAI), 1-1 Oho, Tsukuba, Ibaraki 305-0801, Japan}
\affiliation{Kavli IPMU (WPI), UTIAS, The University of Tokyo, Kashiwa, Chiba 277-8583, Japan}

\date{\today}

\begin{abstract}
 Line-like features in TeV $\gamma$-rays constitute a ``smoking gun'' for TeV-scale particle dark matter and new physics. Probing the Galactic Center region with ground-based Cherenkov telescopes enables the search for TeV spectral features in immediate association with a dense dark matter reservoir at a sensitivity out of reach for satellite $\gamma$-ray detectors, and direct detection and collider experiments. We report on 223~hours of observations of the Galactic Center region with the MAGIC stereoscopic telescope system reaching $\gamma$-ray energies up to 100~TeV. We improved the sensitivity to spectral lines at high energies using large-zenith-angle observations and a novel background modeling method within a maximum-likelihood analysis in the energy domain. No line-like spectral feature is found in our analysis. Therefore, we constrain the cross section for dark matter annihilation into two photons to $\langle \sigma v \rangle \lesssim 5 \times 10^{-28}\,\mathrm{cm^3\,s^{-1}}$ at 1~TeV and $\langle \sigma v \rangle \lesssim 1 \times 10^{-25}\,\mathrm{cm^3\,s^{-1}}$ at 100~TeV, achieving the best limits to date for a dark matter mass above 20~TeV and a cuspy dark matter profile at the Galactic Center. Finally, we use the derived limits for both cuspy and cored dark matter profiles to constrain supersymmetric wino models.
\end{abstract}

\maketitle

\section{Introduction\label{sec:introduction}}

Astrophysical observations suggest the presence of non-baryonic cold dark matter (DM)~\cite{2001Sofue,2006Clowe,2016Planck}. An explanation for DM is the existence of a new class of non-relativistic elementary Weakly Interacting Massive Particles (WIMPs)~\cite{1996Jungman,2010Feng}. WIMPs are supposed to have decoupled from thermal equilibrium in the early Universe and, assuming the DM mass is on the electroweak scale in the GeV to TeV range,  can fully account for the measured relic DM abundance~\cite{1979Steigman}. Pairs  of WIMPs annihilate, producing Standard Model particles including $\gamma$-rays~\cite{1989Bergstrom}. These $\gamma$-ray signals are expected to show distinct spectral features on the energy scale of the WIMP mass, especially if WIMPs directly annihilate into a photon and a second neutral particle like another photon, a $Z$- or $h$-boson~\cite{2012Bringmann}. In such cases, mono-energetic photons are produced at energy $E_{\gamma} = m_{\text{DM}}(1-m^2_{\mathrm{X}}/4m^2_{\text{DM}})$ in the center-of-mass frame, where $m_{\text{DM}}$ and $m_{\mathrm{X}}$ are the rest masses of the DM and second product particle ($\mathrm{X}=\gamma,\,Z,\,h$). In addition,  internal bremsstrahlung  in annihilations into charged Standard Model particles may result in a pronounced $\gamma$-ray spectral feature near the spectral endpoint~\cite{2008Bringmann}. Such sharp spectral features cannot be produced by known astrophysical processes. Hence, if discovered, a TeV $\gamma$-ray line would provide robust evidence for the existence of WIMP DM and new physics. The cross section for direct annihilation into photons is normally loop-suppressed by a factor $\sim \alpha^{2}$, where $\alpha$ is the fine structure constant, compared to competing fermion and gauge boson channels~\cite{1997Bergstrom,2012Bringmann}. However, mechanisms such as Sommerfeld enhancement~\cite{2004Hisano,2009Lattanzi} can significantly increase the signal. For instance, supersymmetric (SUSY) winos on the TeV scale  are predicted to annihilate into $\gamma$-rays with a cross section enhanced by several orders of magnitude compared to the loop-suppressed value~\cite{2004Hisano,2005Hisano,2007Hisano,2009Hisano,2018Beneke,2018Rinchiuso}.

If they exist, $\gamma$-ray signals from DM annihilation will most probably be first seen from nearby DM reservoirs such as the Galactic Center (GC) and dwarf spheroidal galaxies~\cite{1998Bergstroem,2013Doro,2020Rico}. The GC region is particularly promising for such indirect searches for DM because of its high associated DM density. Though, DM searches towards the direction of the GC (SgrA*) are challenged by the extended astrophysical $\gamma$-ray emission detected up to TeV energies in the GC region \cite{2016VERITAS,2018aHESS,2020MAGIC}. $\gamma$-ray telescopes have already extensively searched for DM signatures towards the GC~\cite{2006HESS,2011HESS,2013HESS,2015HESS,2015Fermi,2016HESS,2016aHESS,2017Fermi,2018HESS,2019Fermi}. However, no unambiguous signal has been found so far. The strongest constraint to date on the cross section of WIMP annihilation into two photons in the TeV mass range up to 70~TeV is obtained from 254~hours of observations towards the GC with the H.E.S.S. telescopes~\cite{2018HESS}. In this letter, we present new constraints on $\gamma$-ray lines from WIMP annihilation between  0.9~TeV and 100~TeV from over seven years of observation of the GC region with the MAGIC telescopes, improving previous limits above 20~TeV.

\section{Expected gamma-ray flux\label{sec:gammarayflux}}

The $\gamma$-ray differential flux from  DM annihilation is represented by the equation
\begin{equation}
    \frac{d\Phi_{\gamma}}{dE} = \frac{1}{4\pi} \frac{\langle \sigma v \rangle}{2m^2_{\text{DM}}}\frac{dN_{\gamma}}{dE} \times J(\Delta \Omega)\,,
\label{eq:differential-flux}
\end{equation}
with $\langle \sigma v \rangle$ the thermally averaged cross section of DM annihilation into a $\gamma\gamma$ pair, $m_{\text{DM}}$ the WIMP mass, and $dN_{\gamma}/dE$ the $\gamma$-ray yield per annihilation given by
\begin{equation}
    \frac{dN_{\gamma}}{dE} = 2\delta(E-m_{\text{DM}})\,.
\label{eq:dm-line}
\end{equation}
The so-called $J$-factor,
\begin{equation}
    J(\Delta\Omega) = \int_{\Delta\Omega}d\Omega\int_{\text{l.o.s.}} \rho^2(l, \Omega)\ dl\,,
\label{eq:J-factor}
\end{equation}
is the integral of the squared DM density, $\rho$, over a solid angle $\Delta \Omega$ and along the observed line of sight (l.o.s.), calculated with the \textsc{Clumpy} code~\cite{2019Huetten}. 

The  value of $J$ constitutes the largest uncertainty on the expected annihilation signal from the GC region, as the DM distribution in the central few kpc of the Milky Way is observationally poorly constrained~\cite{2017Iocco,2019Benito,2021Benito}.  Therefore, we consider different DM density models of Milky-Way-like galaxies~\cite{2004Gnedin,2014DiCintio,2016Tollet,2020Lazar}, bracketing the plausible range for the GC in agreement with observations. For the $\Lambda$CDM prediction of a cuspy profile~\cite{2010Navarro,2020Wang}, we adopt the Navarro–Frenk–White (NFW)~\cite{1996Navarro} and Einasto~\cite{2012Retana-Montenegro} models  with the same parameters as in~\cite{2011Pieri,2013HESS,2018HESS}. For the scenario of the inner Galaxy having formed a DM core due to baryonic feedback processes~\cite{2012Pontzen,2012Governato,2014Pontzen,2016Gammaldi}, we discuss the Burkert description~\cite{1995Burkert} from~\cite{2013Nesti} and the fit with a cored Hernquist-Zhao profile~\cite{1990Hernquist,1996Zhao} from~\cite{2017McMillan}, both expressing an almost flat DM density distribution within the Solar circle. Any scenario of smaller cores, as e.g. discussed in~\cite{2015HESS,2021CTA}, or density slopes shallower than the NFW profile results in a $J$-factor bracketed by our profile assumptions. A detailed discussion of the profiles and parameter values is available in the Supplemental Material A provided with this letter~\cite{Supplemental}.

\section{Observations\label{sec:observations}}

We observed the GC with the MAGIC telescopes, located in the Roque de los Muchachos observatory ($28^{\circ}$N, $18^{\circ}$W) at an altitude of 2200~m above sea level on the Canary Island of La Palma in Spain. The MAGIC system consists of two 17 m diameter Imaging Atmospheric Cherenkov Telescopes~\cite{2012MAGIC}. For MAGIC the GC is  visible at zenith angles larger than $58^\circ$. For observations at large zenith angles (LZA), the average distance between the core of the $\gamma$-ray-initiated particle cascade and the telescopes is much larger than for usual observations in the low-zenith range. Therefore, the Cherenkov light pool becomes wider and fewer Cherenkov photons reach the detector due to the increased geometric distance and absorption in the atmosphere. This results in a higher energy threshold for the detection of $\gamma$-rays. At the same time, above the threshold, it enlarges the effective collection area up to an order of magnitude compared to low zenith angles~\cite{2017MAGIC}. Therefore, observing the GC under LZA conditions is well suited for searching line-like $\gamma$-ray emission from TeV-scale DM annihilations. We stress that LZA observations  degrade the energy resolution only by a few percent~\cite{2020Kazuma}.

The MAGIC telescopes have observed the GC region over seven years between April 2013 to August 2020~\cite{2017MAGIC,2020MAGIC}. Observations were conducted in the so-called ``Wobble mode''~\cite{1994Fomin} with different pointing offsets of $0.4^{\circ}$, $0.5^{\circ}$, and $1.0^{\circ}$ with respect to the direction of the GC (SgrA*). The total dataset is divided in nine data subsets, differing in pointing direction of the telescopes and instrumental conditions. We removed low-quality data acquired during sub-optimal observation conditions, e.g. bad weather or strong moonlight and only kept events recorded at zenith angles between $58^{\circ}$ and $70^{\circ}$. After those quality cuts, the total live time of the dataset is 223~hours. In Supplemental Material B~\cite{Supplemental}, we provide details on these data subsets and quality cuts. The data were processed with the MAGIC standard analysis software~\cite{2013Zanin} which uses the Random Forest method~\cite{2008MAGIC} to estimate the energy and arrival direction of the incoming events and to classify the events into $\gamma$-rays and cosmic-ray background.

\section{Data Analysis\label{sec:analysis}}

After the reconstruction, the cut in the particle classification variable was optimized to achieve the best sensitivity for a $\gamma$-ray detection above the hadronic background events. In order to test the data for the existence of a spectral line at the energy $E = m_{\text{DM}}$, we applied the sliding-window analysis technique~\cite{2011Vertongen,2012Weniger}: in a defined energy range around $E$, a global fit to the energy distribution of events is performed within a large circular region of interest (ROI) around the GC by a model composed of the $\gamma$-ray line plus the background from other astrophysical $\gamma$-ray sources and residual charged cosmic rays. The ROI is required to be within a distance  of $1.5^{\circ}$ from the pointing direction to reduce uncertainties from the camera response close to the camera edge, resulting in different ROI sizes for the nine data subsets. The adopted ROIs are given in the Supplemental Material B to this letter~\cite{Supplemental}. For all studied DM profiles, these ROIs were found to provide the best sensitivity to a line search.

The advantage of the sliding-window technique compared to spatial background subtraction methods~\cite{2007Berge} is that it does not require a background sky region off the target (OFF-region). In particular, for spatially extended emission, the signal may leak into an OFF-region, reducing or even removing the sensitivity to the signal. The approach of a sliding-window fit on the energy spectrum does not suffer from this limitation. However, the sliding-window technique is susceptible to a poor background modeling, which we addressed by a careful evaluation of the resulting systematic uncertainties. We searched for a spectral line at 18 different energies, between 0.9~TeV and 100~TeV and set constraints on $\langle \sigma v \rangle$ with the maximum likelihood method. For every data subset $i$ (here $i = 1,...,9$), we modeled the likelihood function as:
\begin{widetext}
\begin{align}
\begin{split}
    \mathcal{L}_i(\langle \sigma v \rangle; \bm{\nu}_i \ | \bm{\mathcal{D}}_i) &= \mathcal{L}_i(\langle \sigma v \rangle; b_{i}, \tau_{i} \ |\{E'_{j}\}_{j=1,...,N_{\text{ON},i}},N_{\text{ON},i}) \\ &= \underbrace{\vphantom{\prod_{j=1}^{N_{\text{ON},i}}}\frac{(g_i+\tau_i b_i)^{N_{\text{ON},i}}}{N_{\text{ON},i} !}e^{-(g_i+\tau_i b_i)}}_{\text{(a)}} \times \underbrace{\prod_{j=1}^{N_{\text{ON},i}}\frac{1}{g_i+\tau_i b_i}(g_i f_{g} (E'_{j}) + \tau_i b_i f_{b} (E'_{j})}_{\text{(b)}})\times \underbrace{\vphantom{\prod_{j=0}^{N_{On}}}\mathcal{T}(\tau_i|\tau_{\text{obs}},\sigma_{\tau})}_{\text{(c)}}\,.
\label{eq:likelihood}
\end{split}
\end{align}
\end{widetext}
In Eq.~\ref{eq:likelihood}, the vector $\bm{\nu}_i$ denotes the nuisance parameters and $\bm{\mathcal{D}}_i$ represents the $i$-th data subset. Term (a) is the Poisson likelihood term for the total number of events observed in the region defined by the energy window around $m_{\text{DM}}$ and the sky region around the pointing direction, with $g_{i}$ and $b_{i}$ the mean number of signal and background events, respectively, and $N_{\text{ON},i}$ the number of observed events in the ROI and energy window. Term (b) is the joint unbinned likelihood for the observed values of the estimated energies, $E'_{j}$. Here, $f_{g}$ and $f_{b}$ are the probability density functions (PDFs) of $E'$ for the signal and background events, respectively. $f_{g}(E') = G(E'|E)$ corresponds to the convolution of the assumed signal spectrum of a sharp spectral line at energy $E$ with the energy dispersion $G$ of the telescopes obtained from Monte Carlo simulations. $f_{b}(E')$ is the assumed background spectral shape in the energy window. The 68\% containment range of the energy resolution, $2\sigma_{E}$, was used to define the energy window, log-centered at $m_{\text{DM}}$ with width $\pm 4 \sigma_{E}$. This width was found to provide the best sensitivity while keeping systematic biases from the background modeling to a minimum. We modeled the combined astrophysical $\gamma$-ray and cosmic-ray backgrounds by interpolating the energy spectrum inside the energy window with a power-law function. Term (c), $\mathcal{T}$, is the likelihood term for the normalization $\tau_i$ of the background, parameterized by a Gaussian function with mean $\tau_{\text{obs}}$ and variance $\sigma^2_{\tau}$. $\tau_{\text{obs}}$ and $\sigma^2_{\tau}$ were estimated from test datasets free of $\gamma$-ray sources, taken off the GC region and in a similar zenith angle range as the dataset on the GC. $\sigma_{\tau}$ includes both statistical and systematic uncertainties and was found to be smaller than 1\% of $\tau_{\text{obs}}$. Details and supporting figures on the determination of $\tau_{\text{obs}}$ and $\sigma_{\tau}$ are provided in the Supplemental Material C to this letter~\cite{Supplemental}. $b_i$ and $\tau_i$ are nuisance parameters, while $g_i$ depends on the free parameter $\langle \sigma v \rangle$ through the following equation:
\begin{multline}
    g_i(\langle \sigma v \rangle) = T_{\text{obs},i} \int_{E'_{\text{min},i}}^{E'_{\text{max},i}} dE' \int_{0}^{\infty} dE \frac{d\Phi(\langle \sigma v \rangle)}{dE}\\
    \times \bar{A}_{\text{eff},i}(E) G_i(E'|E)\,,
\label{eq:mean-number-of-signal-events}
\end{multline}
where $T_{\text{obs},i}$ is the observation time for each data subset $i$, $E$ and $E'$ are the true and the reconstructed energies. $E'_{\text{min},i}$ and $E'_{\text{max},i}$ are the minimum and maximum reconstructed energies in the energy window. The morphology-averaged effective collection area is given by~\cite{2018MAGIC,2020Rico}:
\begin{multline}
    \bar{A}_{\text{eff},i}(E) = \int_{\text{ROI}} d\Omega' \int d\Omega\, \frac{d\mathcal{J}(P)}{d\Omega}\\
    \times A_{\text{eff},i}(E,P)R(P'|E,P)\,.
\label{eq:effective-area}
\end{multline}
Here, $P$ and $P'$ are the true and reconstructed incoming directions corresponding to the differential solid angles $d\Omega$ and $d\Omega'$, respectively. $d\mathcal{J(P)}/d\Omega$ is the PDF for the arrival direction of $\gamma$-rays from DM annihilation around the GC. $A_{\text{eff}}(E, P)$ is the effective collection area for $\gamma$-rays of energy $E$ and arrival direction $P$. $R$ is the PDF of the direction estimator. We computed $\bar{A}_{\text{eff}}$ numerically from a sample of $\gamma$-rays simulated with arrival directions distributed according to $d\mathcal{J(P)}/d\Omega$, applying the so-called ``Donut'' Monte Carlo method~\cite{2018MAGIC}.

The total likelihood $\mathcal{L}$ is the product of the nine individual likelihood terms, $\mathcal{L}_i$, corresponding to the nine considered data subsets:
\begin{equation}
    \mathcal{L}(\langle \sigma v \rangle; \bm{\nu} \ | \bm{\mathcal{D}}) = \prod_{i=1}^9
    \mathcal{L}_i(\langle \sigma v \rangle; \bm{\nu}_i \ | \bm{\mathcal{D}}_i)\,.
\label{eq:total-likelihood}
\end{equation}
Finally, we defined the test statistic (TS):
\begin{equation}
    \text{TS} = -2\mathrm{ln}\lambda_{P}(\langle \sigma v \rangle | \bm{\mathcal{D}}) = -2 \mathrm{ln}\Biggl(\frac{\mathcal{L}( \langle \sigma v \rangle; \hat{\hat{\bm{\nu}}} \ | \bm{\mathcal{D}})}{\mathcal{L}(\langle \widehat{\sigma v}\rangle; \hat{\bm{\nu}} \ | \bm{\mathcal{D}})}\Biggr)\,,
\label{eq:test-statistic}
\end{equation}
where $\langle \widehat{\sigma v } \rangle$ and $\hat{\bm{\nu}}$ in the denominator are the values that maximize $\mathcal{L}$.  In the numerator, $\hat{\hat{\bm{\nu}}}$  is the value maximizing $\mathcal{L}$ for a given $\langle \sigma v \rangle$. The distribution of the TS asymptotically approaches the $\chi^2_{k=1}$ distribution with one degree of freedom according to Wilks' theorem~\cite{2015Conrad}. In the absence of signal, by solving $-2\mathrm{ln}\lambda_{P} = 2.71$, we determined the one-sided 95\% confidence level (CL) to set upper limits on $\langle \sigma v \rangle$.
\footnote{Because of $\langle \sigma v \rangle$ lying at the boundary of the parameter space, the coverage of our confidence intervals is not exactly 95\%. However, by construction (and confirmed using simulations), our recipe produces over-coverage, similarly to other results
~\cite{2015Fermi,2018HESS,2020HAWC,2022MAGIC} derived using similar prescriptions~\cite{2005NIMPA.551..493R}.}

\section{Results and Discussion\label{sec:results}}

\begin{figure}[t!]
\centering
\includegraphics[width=0.48\textwidth]{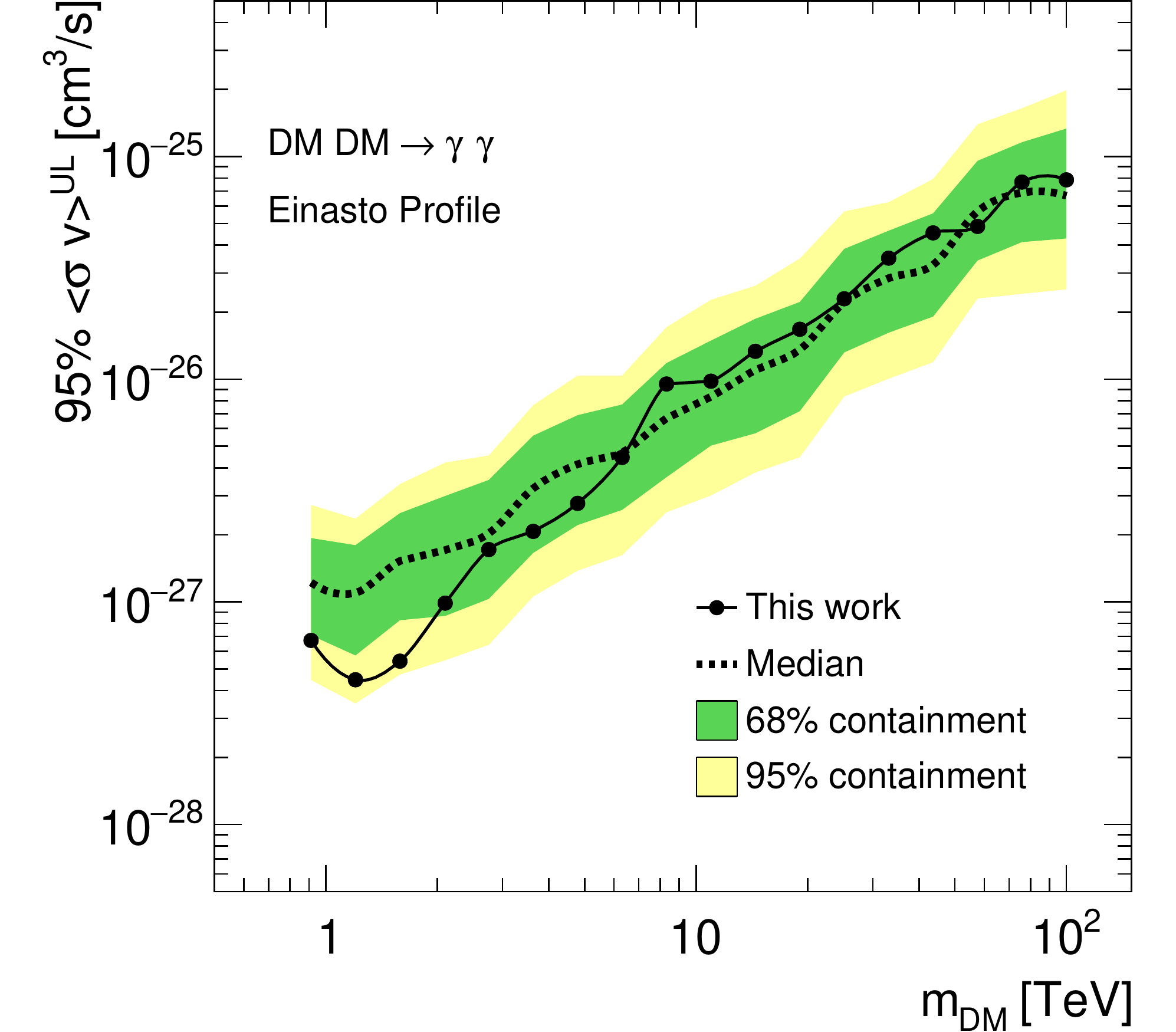}
\caption{95\% CL upper limits on the annihilation cross section $\langle \sigma v \rangle$ into two $\gamma$-rays assuming the Einasto profile, as a function of energy. Observed limits (black dots) are shown together with the mean expected limits (black dotted line) and the 68\% (green) and 95\% (yellow) containment bands.\label{fig:limits_dm_gc}}
\end{figure}

The analysis revealed no significant line-like $\gamma$-ray excess in the GC region. Therefore we derived upper limits on $\langle \sigma v \rangle$ of annihilations into two $\gamma$-rays for DM particle masses between 0.9~TeV and 100~TeV (black dots in Fig.~\ref{fig:limits_dm_gc}, for a cuspy Einasto density profile). We confirmed the consistency of our results with the null hypothesis by performing 300 simulations of the expected background and computing for each tested DM mass the median, the 68\%, and the 95\% containment bands of the obtained distribution of limits on $\langle \sigma v \rangle$ (dotted black curve, green and yellow bands in Fig.~\ref{fig:limits_dm_gc}).

We tested the dependence of our limits on a systematic uncertainty on the energy resolution and a possible bias: we mimicked a detector response with energy resolution of $\sigma_{E}/E = 25\%$, and found our limits to worsen by about 30\%. Correspondingly, a misestimation of the energy scale of 15\% due to unaccounted miscalibration of the telescopes  affects our limits by 30\%.

\begin{figure}[t!]
\centering
\includegraphics[width=0.48\textwidth]{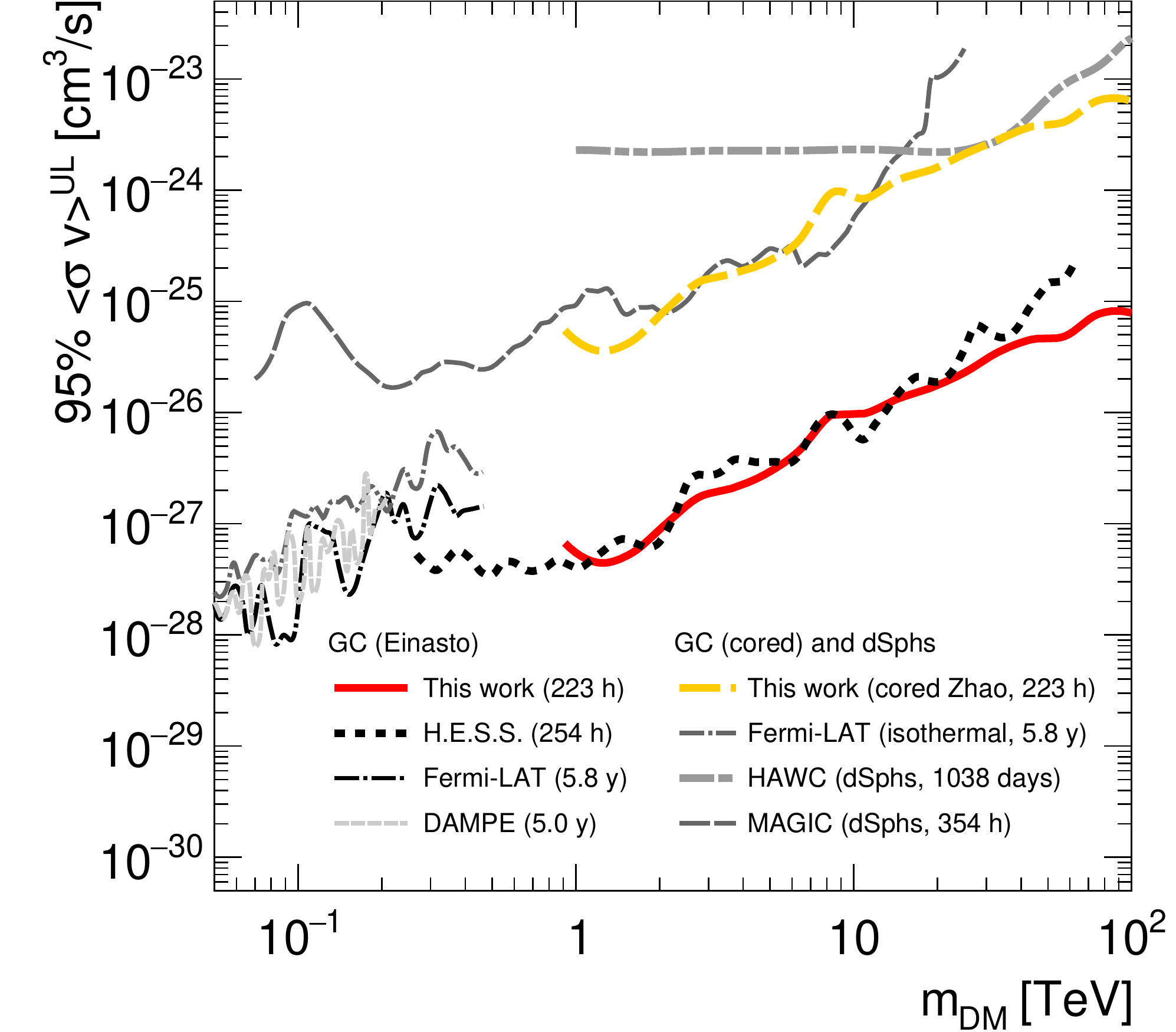}
\caption{95\% CL upper limits to $\gamma$-ray spectral lines from DM annihilation for the Einasto (red solid line) and cored Zhao~(yellow dashed line,~\cite{2017McMillan}) profiles, in comparison to previous works by MAGIC (long gray dashed line,~\cite{2022MAGIC}), \textit{Fermi}-LAT (black and gray dash-dotted lines,~\cite{2015Fermi}), H.E.S.S. (black dotted line,~\cite{2018HESS}), HAWC (gray dash-dotted-dotted line,~\cite{2020HAWC}), and DAMPE (short gray dashed line,~\cite{2021DAMPE}). dSphs: dwarf spheroidal galaxies.\label{fig:limits_comparison_gc}}
\end{figure}

Fig.~\ref{fig:limits_comparison_gc} compares our limits on $\langle \sigma v \rangle$  with previous results by other instruments. The result by the H.E.S.S. telescopes from 2018~\cite{2018HESS} relies on the same Einasto DM halo as in our analysis. Also, the results by \textit{Fermi}-LAT  for 5.8 years of data~\cite{2015Fermi} and recently by DAMPE for 5 years of data~\cite{2021DAMPE} are given for an almost identical Einasto halo as in our work. Our result using the Einasto profile is competitive with the current best limits in the mass range of a few TeV and improve the best limits above 20~TeV by a factor of 1.5 to 2. This improvement in sensitivity is due to increased statistics at TeV energies by LZA observations. Fig.~\ref{fig:limits_comparison_gc} also shows how the uncertain knowledge about the DM distribution in the inner Galaxy~\cite{2013Nesti,2017McMillan} impacts our limits. In case of an extended DM core around the GC, our constraints on $\langle \sigma v \rangle$ worsen by about two orders of magnitude. This degradation is caused by the shallower profile shape resulting in a lower $J$-factor in the ROI. We emphasize that our analysis allows to derive limits for such cored profiles, which is challenging for spatial background subtraction methods, as applied by e.g.~\cite{2016aHESS,2018HESS}. Our conservative limits on $\langle \sigma v \rangle$, corresponding to the lowest DM density in the inner Galaxy compatible with observational data, are comparable to the current most stringent limits from observation of dwarf galaxies, as shown for MAGIC~\cite{2022MAGIC} by the gray dashed curve in Fig.~\ref{fig:limits_comparison_gc}.

\begin{figure}[t!]
\centering
\includegraphics[width=0.48\textwidth]{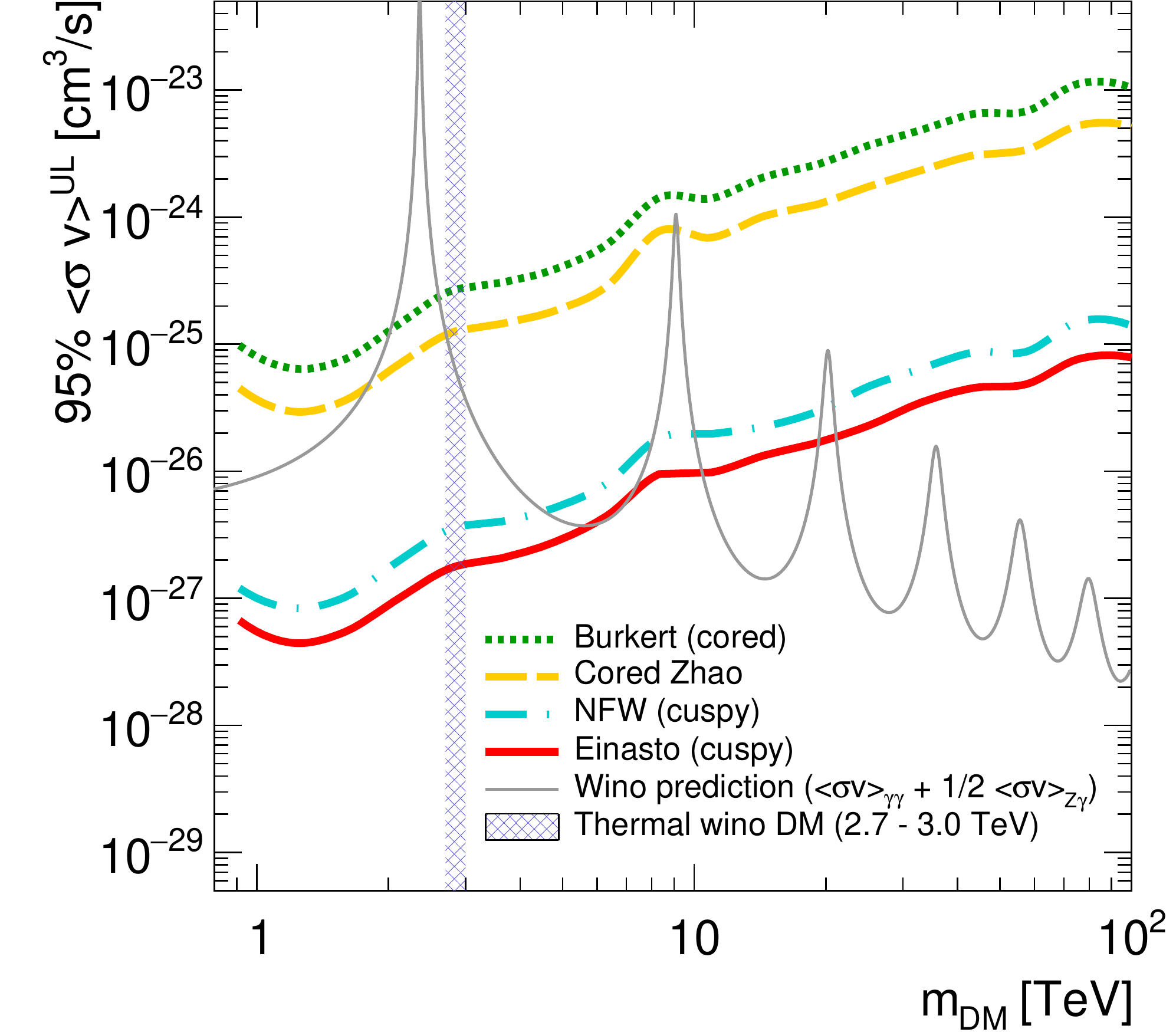}
\caption{Upper limits for the four DM density profiles considered in this work: the cuspy Einasto Galactic density profile (red solid line), the NFW profile (cyan dashed line), a DM core according to~\cite{2017McMillan} (yellow dashed line) and the Burkert fit from~\cite{2013Nesti} (green dotted line), compared against the total $\langle \sigma v \rangle$ corresponding to annihilation  of two SUSY winos (i.e., SU(2)$_L$ triplets) into a $\gamma\gamma$ pair according to~\cite{2004Hisano,2005Hisano,2007Hisano, 2009Hisano} (gray solid line, see text for details). The vertical blue hatched region indicates wino masses from 2.7 to 3.0~TeV which are consistent with the observed DM relic density~\cite{2007Hisano}.\label{fig:limits_susy_wino_gc}}
\end{figure}

Our upper limits are able to constrain heavy SUSY models for both cuspy and cored profiles. In Fig.~\ref{fig:limits_susy_wino_gc}, we show our limits for the two cuspy and two cored profiles introduced in Sec.~\ref{sec:gammarayflux} compared to the total cross section of the two annihilation processes\footnote{The factor 1/2 for the $Z\gamma$ channel expresses that in the calculation of our limits we have assumed the production of two $\gamma$-rays per annihilation process (Eq.~\ref{eq:dm-line}), whereas for this channel only one is produced.} into $\gamma\gamma$ and $Z\gamma$ pairs for the wino model from~\cite{2004Hisano,2005Hisano,2007Hisano,2009Hisano}. The resonances in the thin gray curve show the Sommerfeld enhancement of the branching ratio and overall annihilation cross section for winos of the respective masses. Therefore, for the cuspy profiles, we can exclude wino annihilations for masses below 5~TeV and especially in the range between 2.7 and 3.0~TeV, found to produce a consistent thermal relic DM abundance~\cite{2007Hisano} (blue hatched band in Fig.~\ref{fig:limits_susy_wino_gc}). In turn, for the most conservative assumptions about a cored halo profile, a 2.7~TeV wino would be just marginally in agreement with our null measurement.

\section{Conclusion and Summary\label{sec:conclusion}}

We have presented a search for spectral lines in $\gamma$-rays from 0.9~TeV to 100~TeV towards the Galactic Center using 223~hours of observations with the MAGIC telescopes. The sensitivity at these high energies is boosted by the large telescope acceptance in LZA observations. In the analysis, we have used a sliding-window technique in the energy domain to search for a line-like signal on the top of the astrophysical $\gamma$-ray and cosmic-ray backgrounds. This approach has provided us with an unprecedented sensitivity to search for a signal from either a localized or very extended region in the sky. With this, we could probe the GC region for emission from DM annihilation for both the optimistic and conservative assumptions of a cuspy or cored Galactic DM halo. We have not found a significant signal of line-like $\gamma$-ray emission and have computed upper limits  on the WIMP annihilation cross section $\langle \sigma v \rangle$. Around 1~TeV, the observed upper limits reach $\simeq 5 \times 10^{-28}\,\mathrm{cm^3\,s^{-1}}$ with the Einasto profile and $\simeq 8 \times 10^{-26}\,\mathrm{cm^3\,s^{-1}}$ with the Burkert profile. At 100~TeV the limits reach below $1 \times 10^{-25}\,\mathrm{cm^3\,s^{-1}}$ in the Einasto case and $\simeq 1 \times 10^{-23}\,\mathrm{cm^3\,s^{-1}}$ in the Burkert case. This represents competitive limits on the line-like annihilation of TeV DM into $\gamma$-rays, with up to a factor 2 better sensitivity above 20~TeV compared to previous measurements, and for the first time probed up to 100~TeV with Imaging Air Cherenkov Telescopes.

\begin{acknowledgments}
We would like to thank the Instituto de Astrof\'{\i}sica de Canarias for the excellent working conditions at the Observatorio del Roque de los Muchachos in La Palma. The financial support of the German BMBF, MPG and HGF; the Italian INFN and INAF; the Swiss National Fund SNF; the grants PID2019-104114RB-C31, PID2019-104114RB-C32, PID2019-104114RB-C33, PID2019-105510GB-C31, PID2019-107847RB-C41, PID2019-107847RB-C42, PID2019-107847RB-C44, PID2019-107988GB-C22 funded by MCIN/AEI/ 10.13039/501100011033; the Indian Department of Atomic Energy; the Japanese ICRR, the University of Tokyo, JSPS, and MEXT; the Bulgarian Ministry of Education and Science, National RI Roadmap Project DO1-400/18.12.2020 and the Academy of Finland grant nr. 320045 is gratefully acknowledged. This work was also been supported by Centros de Excelencia ``Severo Ochoa'' y Unidades ``Mar\'{\i}a de Maeztu'' program of the MCIN/AEI/ 10.13039/501100011033 (SEV-2016-0588, SEV-2017-0709, CEX2019-000920-S, CEX2019-000918-M, MDM-2015-0509-18-2) and by the CERCA institution of the Generalitat de Catalunya; by the Croatian Science Foundation (HrZZ) Project IP-2016-06-9782 and the University of Rijeka Project uniri-prirod-18-48; by the DFG Collaborative Research Centers SFB1491 and SFB876/C3; the Polish Ministry Of Education and Science grant No. 2021/WK/08; and by the Brazilian MCTIC, CNPq and FAPERJ. This work was supported by JSPS Grant-in-Aid for JSPS Research Fellow 19J12715. This project has received funding from the European Union's Horizon 2020 research and innovation programme under the Marie Sk\l{}odowska-Curie grant agreement No. 754510.
\end{acknowledgments}

\textbf{Author Contributions:} Nagisa Hiroshima: theoretical interpretation. Moritz Hütten: astrophysical modeling, theoretical interpretation, paper drafting, writing, and edition. Tomohiro Inada: project leadership, MAGIC data analysis, software development for the likelihood analysis, theoretical interpretation, paper drafting and edition. Daniel Kerszberg: MAGIC data analysis cross-check, software development for the likelihood analysis, theoretical interpretation, paper drafting and edition. Kazunori Kohri: theoretical modeling of particle physics, theoretical interpretation, providing theoretical data. All other authors have contributed in one or several of the following ways: design, construction, maintenance and operation of the instrument used to acquire the data; preparation and/or evaluation of the observation proposals; data acquisition, processing, calibration and/or reduction; production of analysis tools and/or related Monte Carlo simulations; discussion and approval of the contents of the draft.

\bibliography{main}

%%%%for arXivsubmission
\clearpage

\onecolumngrid
\setcounter{page}{1}
\setcounter{figure}{0}
\setcounter{equation}{0}

\noindent\rule{17.5cm}{0.3pt}
\section*{Supplemental Material}
\begin{center} 
Search for Gamma-ray Spectral Lines from Dark Matter Annihilation up to 100 TeV towards the Galactic Center with MAGIC (The MAGIC Collaboration)
\end{center}
\noindent\rule{17.5cm}{0.3pt}

\subsection{Model choices and parameters for the Milky Way dark matter density distribution\label{subsec:dm-densities}}

In this work, we have described the Milky Way (MW) dark matter (DM) halo with analytic prescriptions commonly adopted in the literature, with parameters fit either to simulation results or kinematic data. For the different models, we have made use of the Einasto profile~\cite{2012Retana-Montenegro}:
\begin{equation}
    \rho_{\text{Einasto}}(r) = \rho_s \exp\left\{\cfrac{-2}{\alpha} \left[\left(\frac{r}{r_s}\right)^\alpha -1 \right] \right\}\,,
\label{eq:einasto}
\end{equation}
the Hernquist-Zhao $(\alpha,\beta,\gamma)$ profile~\cite{1990Hernquist,1996Zhao}:
\begin{equation}
    \rho_{\text{Zhao}}(r) = \cfrac{2^{\frac{\beta-\gamma}{\alpha}}\rho_s}{\left(\cfrac{r}{r_s}\right)^\gamma \left[ 1 + \left( \cfrac{r}{r_s} \right)^{\alpha}\right]^{\frac{\beta-\gamma}{\alpha}}}\,,
\label{eq:nfw}
\end{equation}
as well as the cored Burkert model~\cite{1995Burkert}:
\begin{equation}
    \rho_{\text{Burkert}}(r) = \cfrac{\rho_s}{\left(1+\cfrac{r}{r_s}\right)\left(1+\cfrac{r^2}{r_s^2}\right)}\,.
\label{eq:burkert}
\end{equation}
For the two considered cases describing  a steep DM density cusp in the inner Galactic halo, we have adopted the Einasto profile with $\alpha=0.17$ and Navarro-Frenk-White (NFW, Eq.~(\ref{eq:nfw}) with $\alpha=1,\,\beta=3,\,\gamma=1$) descriptions from~\cite{2011Pieri}, but with slightly modified values of $\rho_s$ as used in~\cite{2013HESS,2018HESS}, calibrating the local DM density $\rho_\odot$ (and consequently, the halo mass) to a marginally lower value. Although recent analyses using data from the GRAVITY experiment and GAIA satellite suggest a somewhat higher value in the range of $0.4-0.8\,\mathrm{GeV\,cm^{-3}}$~\cite{2019Benito,2021Benito}, the chosen values are still in the allowed observational range, and we have kept the profiles identical to~\cite{2013HESS,2018HESS} to ease comparison of the results.

\begin{figure}
\centering
\includegraphics[width=0.8\columnwidth]{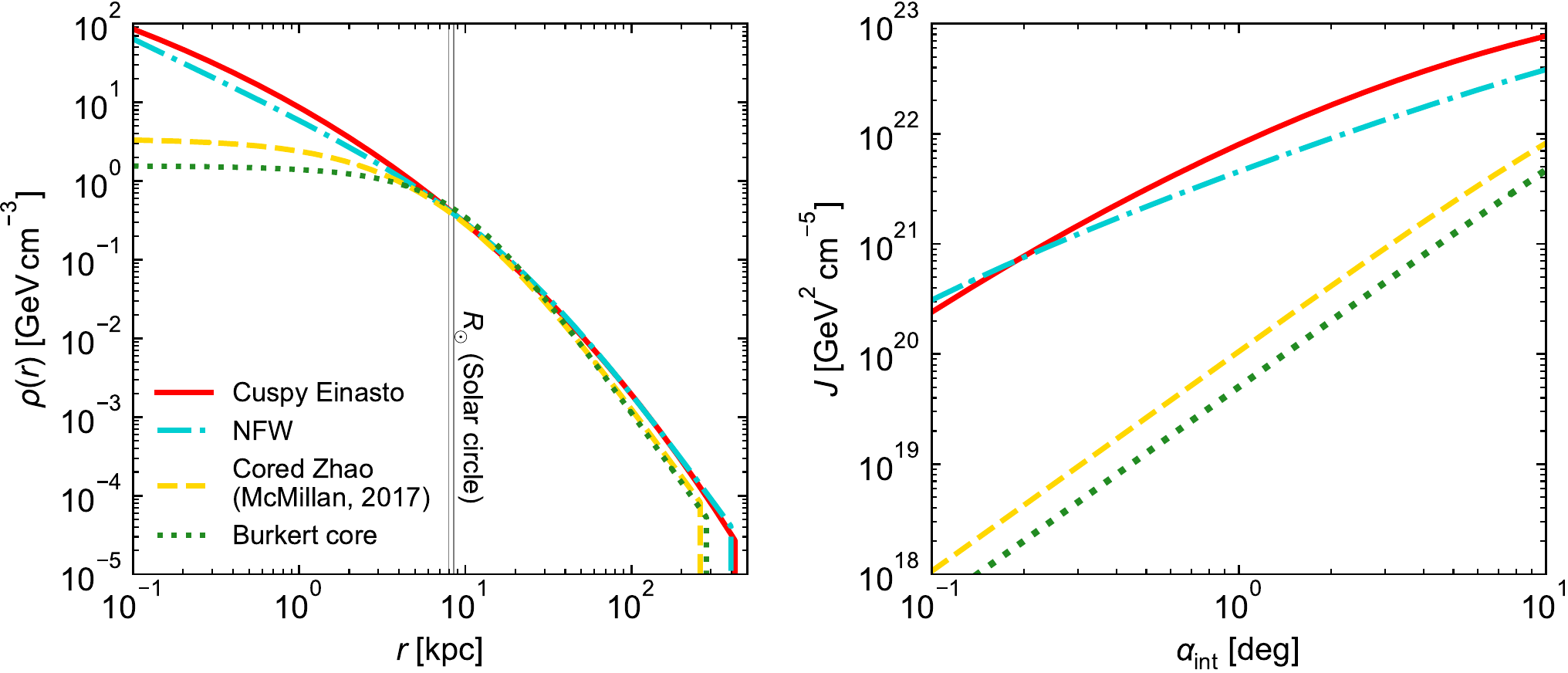}
\caption{Comparison of the different density models and resulting $J$-factors for the spherical MW DM halo considered in this work. Left: spherical density profiles. Right: corresponding $J$-factors within circular integration regions centered at the GC and radius $\alpha_{\text{int}}$.\label{fig:dm-densities-int}}
\end{figure}

While standard $\Lambda$CDM cosmology predicts scale-invariant cuspy density profiles of DM halos~\cite{2020Wang}, the presence of baryons can significantly alter the inner cusps of DM halos. The impact of baryonic physics onto the DM profile is complex with counter-acting processes: the DM density is expected to flatten by non-adiabatic feedback by star formation and supernova winds~\cite{2012Pontzen}, while in turn baryonic energy and angular momentum dissipation contract the DM profile~\cite{2004Gnedin}. Different processes dominate on different galaxy mass scales, with no clear trend for the behavior of a MW-sized galaxy~\cite{2014DiCintio,2016Tollet,2020Lazar}. While cores not larger than $0.5-1\,\mathrm{kpc}$ in radius are generally expected to form in MW-like galaxies~\cite{2012Governato,2014Pontzen,2016Gammaldi}, a core as large as several kpc in radius is not excluded observationally for the MW, and is preferred in fits with a fixed flat DM distribution at the Galactic Center (GC) as the Burkert profile from~\cite{2013Nesti} and Hernquist-Zhao profile with $\alpha=1,\,\beta=3,\,\gamma=0$ from~\cite{2017McMillan} investigated in this paper. We remark that even the less conservative case~\cite{2017McMillan} of our two considered core models provides a DM density in the inner GC region a factor~3 smaller than an Einasto halo with a flat core within 1 kpc in radius, and even a factor 10 smaller densities compared to such Einasto profile cored at 0.5 kpc (see Fig.~\ref{fig:dm-densities-int}, left).

In Tab.~\ref{tab:density-values}, we list the choices of parameters adopted for the different density profile models. We provide the density normalization both in terms of the profile scale radius, $\rho_s = \rho(r_s)$, as well as the local DM density at the Solar circle, $\rho_\odot = \rho(R_\odot)$. $R_{\text{max}}$ denotes, as a measure of the MW virial radius, the maximum radius at which we stopped the line-of-sight integration according to Eq.~(3) of the main paper. 

In Tab.~\ref{tab:J-factor-values}, we provide the $J$-factor values integrated according to Eq.~(3) of the main paper within a radius corresponding to the one of the region of interests (ROIs, see Fig.~\ref{fig:pointing}) centered on the GC. In Fig.~\ref{fig:dm-densities-int}, we compare the different adopted density models (left) and resulting $J$-factors as a function of angular distance from the GC (right).

\begin{table}[ht]
\centering
\begin{tabular}{lcccccccccc} \hline\hline
Profile name & Profile type & $\alpha$ & $\beta$ & $\;\;\gamma$ & $\;\;\rho_s$ $\mathrm{[GeV\,cm^{-3}]}$ & $r_s$ $\mathrm{[kpc]}$ & $\rho_{\odot}$ $\mathrm{[GeV\,cm^{-3}]}$ & $R_{\odot}$ $\mathrm{[kpc]}$ & $R_{\text{max}}$ $\mathrm{[kpc]}$ & Reference \\
\hline
Cuspy Einasto & Einasto & 0.17 & -- & $\;\;$-- & 0.0790 & 20 & 0.388 & 8.5 & 433 & \cite{2011Pieri,2013HESS,2018HESS} \\
NFW & Zhao & 1 & 3 & $\;\;$1 & 0.0768 & 21 & 0.384 & 8.5 & 402 &  \cite{2011Pieri,2013HESS,2018HESS} \\
Cored Zhao  & Zhao & 1 & 3 & $\;\;$0 & 0.431 & 7.7 & 0.391 & 8.21 & 265 & \cite{2017McMillan} \\
Burkert core & Burkert & -- & -- & $\;\;$-- & 1.568 & 9.26 & 0.487 & 7.94 & 291 & \cite{2013Nesti} \\
\hline
\end{tabular}
\caption{Parameter choices and selected halo properties of the Galactic DM density models considered in this work. Note the different definition of $\rho_s$ in the Hernquist-Zhao profile of Eq.~(\ref{eq:nfw}). when comparing against~\cite{2013HESS,2017McMillan,2018HESS}. For the Einasto and NFW profiles, $R_{\text{max}}$ is taken from~\cite{2011Pieri}. For the Hernquist-Zhao and Burkert core profiles, we chose $R_{\text{max}}$ such that to obtain the halo masses given in~\cite{2013Nesti} and~\cite{2017McMillan}.\label{tab:density-values}}
\end{table}

\begin{table}[ht]
\centering
\begin{tabular}{lccc} \hline\hline
Profile name            & $J(0.5^\circ)$            & $J(1.0^\circ)$ & $J(1.1^\circ)$\\
\hline
Cuspy Einasto           & $3.14\times10^{21}$       & $8.01\times10^{21}$      & $9.03\times10^{21}$\\
NFW                     & $2.18\times10^{21}$       & $4.55\times10^{21}$      & $5.02\times10^{21}$\\
Cored Zhao   & $2.66\times10^{19}$       & $1.06\times10^{20}$      & $1.28\times10^{20}$\\
Burkert core            & $1.26\times10^{19}$       & $5.04\times10^{19}$      & $6.10\times10^{19}$\\
\hline
\end{tabular}
\caption{$J$ values integrated within a radius corresponding to the one of the ROIs, i.e. $0.5^{\circ}$, $1.0^{\circ}$, and $1.1^{\circ}$, around the direction towards the GC (see Fig.~\ref{fig:pointing}) and models from Tab.~\ref{tab:density-values}. All values are given in units of $\mathrm{GeV^2\,cm^{-5}}$.\label{tab:J-factor-values}}
\end{table}

\subsection{Observational dataset and definition of the region of interest\label{subsec:dataset}}
 
\begin{figure}
    \centering
    \includegraphics[width=188mm]{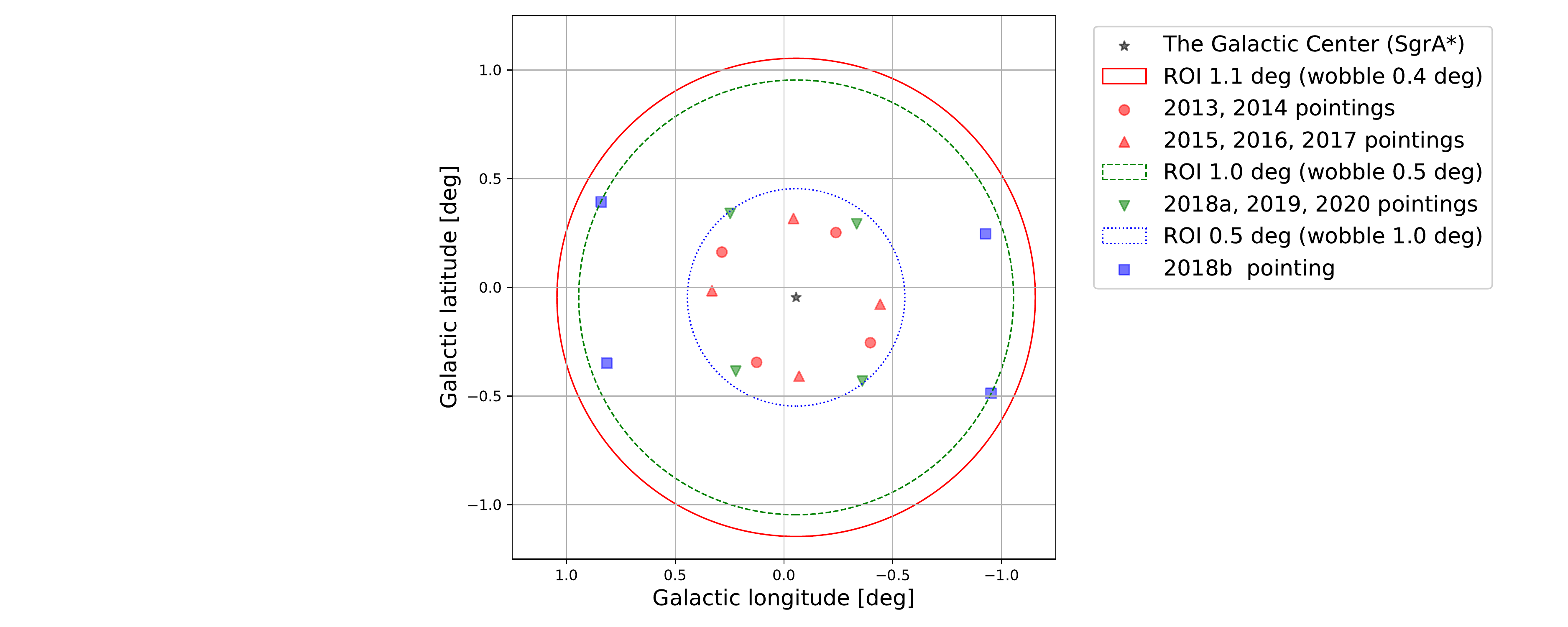}
    \caption{Pointing positions (Galactic coordinates) of the MAGIC telescopes in the used data subsets and adopted ROIs. The markers show the pointing directions. The position of the GC is displayed with a black star, around which the ROIs of our analysis are centered for all data subsets. The ROI encircled by the red solid line has a radius of $1.1^\circ$ and is used for the data from 2013 and 2014 with the telescope pointings marked by the red dots, and for and 2015 and 2016 with the pointings marked by red upright triangles. The ROI indicated by the green dashed circle has a radius of $1.0^\circ$ and is used for the data from 2018 to 2019 using the pointings marked by the green upside down triangles. The blue dotted solid circle shows the ROI with $0.5^\circ$ radius for the data taken in 2018 with the pointings marked by the blue squares.\label{fig:pointing}}
\end{figure}

The GC has been observed over many years by MAGIC. During these years, several upgrades of the instrument have been deployed and different pointing directions were used. Therefore, our dataset is divided into nine subsets of constant instrumental conditions and pointing offsets, listed in Tab.~\ref{tab:dataset}. For all subsets, data was taken with the telescopes pointing slightly offset from the GC (SgrA*), with different offsets and different directions. Fig.~\ref{fig:pointing} illustrates the various pointing directions in Galactic coordinates around the GC. Our ROIs are circular regions of $0.5^{\circ}$, $1.0^{\circ}$, and $1.1^{\circ}$ in radius around the GC position, depending of the telescopes' pointing direction in the data subsets, and are marked by the respective circles.  Note that by this configuration, the ROIs are located at different positions in the telescopes' field of view, depending on the pointing. By these ROI choices, the maximum distance of a reconstructed event from the camera center used in the analysis is kept below $1.5^\circ$ for all data subsets.

The total observation time reached about 260~hours. We applied various selection cuts to ensure the data quality. Main selection cuts were based on (1) the atmospheric transmission, (2) the night-sky background, and (3) the shower image quality. For (1), a LIDAR was measuring the differential transmission of the atmosphere during the observations. We removed from the dataset time periods with less than 80~\% atmospheric transparency. For (2), the direct current of the photomultiplier tubes in the camera reflects the sky brightness, also an indicator of the weather conditions: when clouds appear in the sky, they reflect light from the ground and the sky brightness increases. The typical value for a direct current cut is requiring less than 1.4~$\mu$A for MAGIC-I and 3.0~$\mu$A for MAGIC-II during astronomical dark time. For (3), a cut on the minimal total charge contained in a shower image is applied. We removed events that have less than 50~photoelectrons after the image cleaning procedure to efficiently suppress misreconstruction of the total charge by night sky background pollution. The night sky background rate is typically 100 - 120 MHz/pixel which corresponds to about $\sim 0.17$~photoelectron/pixel/ns.

\begin{table}[ht]
\centering
\begin{tabular}{cccc} \hline
\multirow{2}{*}{Dates} & \multirow{2}{*}{Label}& Total observation time [h] & Effective live time [h] \\%& Pointing offset [deg]\\
 & & (before quality cuts) & (after quality cuts) \\
\hline
2013/03/10 -- 2013/07/18 & 2013 & 47.1 & 38.8 \\%& 0.40\\
2014/03/01 -- 2014/07/07 & 2014 & 37.3 & 30.1 \\%& 0.40\\
2015/03/29 -- 2016/04/13 & 2015 & 27.0 & 18.9 \\%& 0.40\\
2016/05/02 -- 2016/08/05 & 2016 & 24.8 & 17.3 \\%& 0.40\\
2017/03/26 -- 2017/06/24 & 2017 & 26.0 & 22.1 \\%& 0.40\\
\multirow{2}{*}{2018/02/19 -- 2018/09/30} & 2018a &  26.3 & 19.1 \\%& 0.50 \\
 & 2018b & 7.0& 5.8 \\%&1.00 \\
2019/03/11 -- 2019/08/04 & 2019 & 54.4& 52.0 \\%&0.50\\
2020/06/19 -- 2020/08/21 & 2020 & 22.9 & 19.1 \\%&0.50\\
\hline  
Total & & 272.8 & 223.2 \\
\hline
\end{tabular}
    \caption{Observational periods of constant instrumental conditions (specified by the time ranges) and pointing directio    ns (see Fig.~\ref{fig:pointing} and the label column), with their corresponding raw observation times and effective (i.e. dead time corrected) live times after quality cuts.\label{tab:dataset}}
\end{table}

\subsection{Determination of the background normalisation parameter \texorpdfstring{$\tau_{\text{obs}}$}{tau} and its variance\label{subsec:tau}}

\begin{figure}
\centering
\includegraphics[width=0.8\columnwidth]{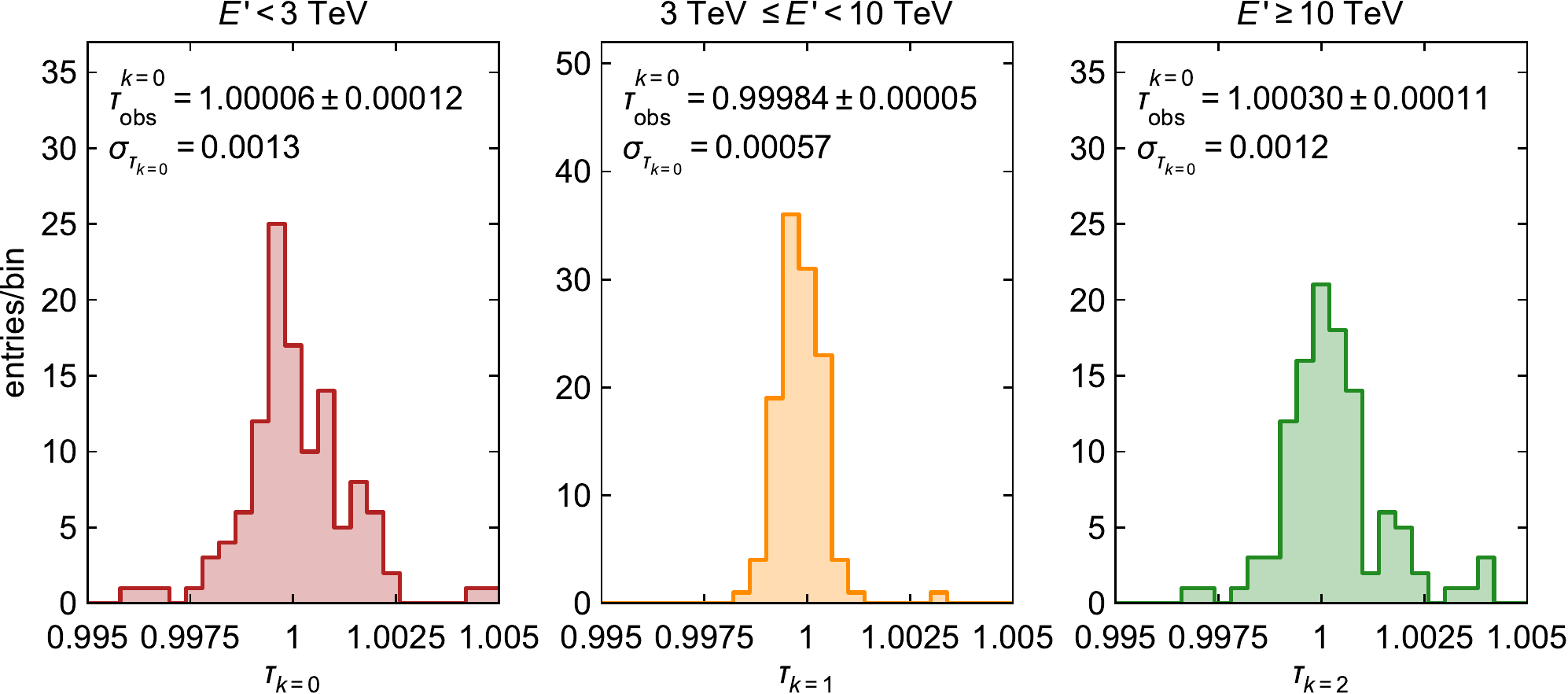}
\caption{Distributions of $\tau$ according to Eq.~(\ref{eq:tau}) when searching for line signals in the test data in the three energy intervals, based on 120 analyzed test datasets in each interval. It is found  a small bias $\tau_{\text{obs}} < 1$ for energies $E' \gtrsim$ 3 TeV and $\sigma_{\tau^k}< 0.01\,\tau_{\text{obs}}^k$. The statistical error according to Eq.~(\ref{eq:sigma-tau-stat}) is $\sigma_{\tau=0, \text{stat}}=2.60\times 10^{-5}$, $\sigma_{\tau=1, \text{stat}}=2.76\times 10^{-5}$, and $\sigma_{\tau=2, \text{stat}}=1.77\times 10^{-4}$. \label{fig:tau_dist}}
\end{figure}

To determine the likelihood parameters $\tau_{\text{obs}}$ and $\sigma_{\tau}$, we chose suitable test datasets far off the GC and free of any known $\gamma$-ray signals, but in the same range of zenith angles as the GC dataset. We performed on these test data the same analysis as for the GC data according to Eq.~(4) of the main paper. The analysis was applied on 20 independent test datasets, all with a ROI radius of $1.1^\circ$, for each of the 18 probed DM masses, resulting in 360 samples in total. Then, for each of the samples, the quantity $\tau$ given by
\begin{equation}
    \tau = \frac{N_{\text{ON}}-N_{\text{sig}}}{N_{\text{ON}}}\,
\label{eq:tau}
\end{equation}
was computed. The resulting distribution has the mean $\tau_{\text{obs}}$ and statistical variance
\begin{equation}
    \sigma_{\tau, \text{stat}}^{\;2} = \left(\frac{\partial \tau}{\partial N_{\text{sig}}} \times \sigma_{N_{\text{sig}}} \right)^2 + \left(\frac{\partial \tau}{\partial N_{\text{ON}}} \times \sigma_{N_{\text{ON}}}\right)^2\,,
\label{eq:sigma-tau-stat}
\end{equation}
where $N_{\text{ON}}$ is the number of observed events in the sliding window, and $N_{\text{sig}}$ the number of events associated to a fitted signal component. This procedure allowed us to consider a possible bias in $\tau_{\text{obs}}$, i.e. $\tau_{\text{obs}} \neq 1$. Furthermore, to take into account a potential energy dependence of $\tau_{\text{obs}}$ and $\sigma_\tau$, we calculated $\tau^k_{\text{obs}}$ and $\sigma_{\tau^k}$ in three energy intervals ($k=0, 1, 2$), namely, for  $E' < 3\,\text{TeV}$, $3\,\text{TeV} \leq E' < 10\,\text{TeV}$, and $E' \geq 10\,\text{TeV}$, where in each interval we determined $\tau_{\text{obs}}^k$ and of $\sigma_{\tau^k}$ from 120 samples by merging the analyses of six masses to increase the statistical power of the result. The observed total variance of $\tau$ in each energy interval can be written as
\begin{equation}
    \sigma_{\tau^k}^{\;2} = \sigma_{\tau^k, \text{stat}}^{\;2} + \sigma_{\tau^k, \text{syst}}^{\;2}\,.
\label{eq:sigma-tau-total}
\end{equation}
If $\sigma_{\tau^k}$ was only driven by the Poissonian fluctuations of the number of events, the variance would match the expected statistical uncertainty $\sigma_{\tau^k, \text{stat}}$ according to Eq.~(\ref{eq:sigma-tau-stat}). However, we found $\sigma_{\tau^k}$ to be dominated by $\sigma_{\tau^k, \text{syst}}$, and obtained $\sigma_{\tau^k}\approx\sigma_{\tau^k, \text{syst}} < 0.01\,\tau_{\text{obs}}^k$ (see Fig.~\ref{fig:tau_dist}). It can be seen in Fig.~\ref{fig:tau_dist} a small bias $\tau_{\text{obs}} < 1$ for energies $E' \gtrsim$ 3 TeV, attributed to the approximation of the background spectral shape by a power law in the sliding window. On the other hand, the width of $\sigma_{\tau^k}$, taken into account in the likelihood fitting (see term (c) of Eq.~(4) in the main paper), indicates a variation of the expected background spectral shape rendering this bias negligible.

\end{document}